\documentclass{appolb}
\usepackage{graphicx}
\usepackage{amssymb}
\usepackage{graphicx}
\usepackage{amsmath}
\usepackage{graphicx}
\usepackage{dcolumn}
\usepackage{bm}
\usepackage{epsfig}
\usepackage{color}
\usepackage{datetime}
\usepackage[T1]{fontenc}
\usepackage[utf8]{inputenc}

\newcommand{\bal}{\begin{align}}
\newcommand{\eal}{\end{align}}
\newcommand{\beq}{\begin{eqnarray}}
\newcommand{\eeq}{\end{eqnarray}}
\newcommand{\nneeq}{\nonumber \end{eqnarray}}

\newcommand{\nn}{\nonumber \\}
\newcommand{\es}{& = &}
\newcommand{\rs}{\, = \,}
\newcommand{\ps}{& + &}
\newcommand{\ms}{& - &}
\newcommand{\ts}{& \times &}

\newcommand{\nt}{\nn \ts}
\newcommand{\np}{\nn \ps}
\newcommand{\nm}{\nn \ms}

\newcommand{\tl}{ & \to & }

\newcommand{\cM}{ {\cal M} }
\newcommand{\cH}{ {\cal H} }

\newcommand{\cV}{ {\cal V} }

\newcommand{\cL}{ {\cal L} }

\newcommand{\2}{ \, { 1 \over 2} \, }

\begin{document}
\title{ Massive Abelian gauge bosons in front-form Hamiltonians }
\author{ Stanis{\l}aw D. G{\l}azek 
\address{ Institute of Theoretical Physics,
              Faculty of Physics, 
              University of Warsaw}}
\date{ November 23, 2018 }
\maketitle
\begin{abstract}
It is shown how gauge bosons can be supplied a mass term 
using the Higgs mechanism for the purpose of regulating 
Hamiltonians of Abelian gauge theories in the front form of 
quantum dynamics.  
\end{abstract} 

\section{ Introduction }
\label{intro}

Introduction of a mass parameter for gauge bosons that is
discussed in this article is motivated by appearance of infrared 
(IR) divergences in quantum Hamiltonians of the gauge theories, 
in which gauge bosons are massless. The canonical Hamiltonian 
of QCD in the front form (FF) of dynamics~\cite{DiracFF} 
provides an important example of such divergences~\cite{Wilsonetal}. 
Using the parton model~\cite{Feynman} language of the 
infinite momentum frame (IMF)~\cite{Weinberg}, one can 
say that each of the quarks and gluons in a hadron carries 
some fraction $x$ of the hadron momentum and some momentum 
$k^\perp$ that is spatially orthogonal to the hadron momentum. 
The Hamiltonian contains functions of $x$ that diverge when $x 
\to 0$. In the evolution of quark and gluon states, one encounters 
the sums over intermediate states that involve integrals of the 
type $\int dx/x$ and $\int d^2k^\perp/k^{\perp \, 2}$. These 
integrals produce IR infinities due to small $x$ and $k^\perp$. 
Literature on the FF of Hamiltonian dynamics in quantum field 
theory (QFT) is reviewed in Ref.~\cite{FFreview}.  Examples of 
application to the standard model are offered in 
Refs.~\cite{LepageBrodsky,Srivastava}. 

The issue is that separate regularizations of small $x$ and 
$k^\perp$ regions in phase space introduce the frame 
dependence that is difficult to remove from the theory. 
Ultraviolet (UV) divergences in $k^\perp$ complicate the 
situation because the  unknown finite parts of the UV 
counter-terms interfere with the small-$x$ dynamics; they 
introduce unknown functions of $x$~\cite{Wilsonetal}. 
Examples of functions of $x$ one may expect in the case of 
asymptotically free theories are described in Ref.~\cite{Gomez}. 
Reference~\cite{Gomez} employs the renormalization group
procedure for effective particles (RGPEP)~\cite{RGPEP}.
The regularization of IR singularities due to massless gauge
bosons by introduction of a mass parameter for them is 
discussed here for the purpose of application in the RGPEP.

The complex frame dependence in regularization of FF 
Hamiltonians may be mitigated by introducing a mass 
parameter for gauge bosons because a mass does not 
depend on a frame of reference. However, the gauge 
boson mass parameter introduced arbitrarily leads to 
severe complications in the RGPEP calculation of counterterms.
Knowing that the Higgs mechanism~\cite{Higgs1,Higgs2,
Higgs3,EnglertBrout} leads to a renormalizable theory in 
the path-integral formulation of QFT with a finite number
of counter-terms, one may take advantage of this mechanism 
in the RGPEP for FF Hamiltonians. This article concerns the 
issue of how to introduce the mass as a regularization 
parameter in the case of Abelian theory.

A comment is in order regarding small-$x$ singularities
due to fermions. Since fermions are considered to have 
non-zero masses, their mass parameters can be directly
used for the purpose of regularization.

An additional reason of interest in introducing a mass
parameter for gauge bosons is that the limit $x \to 0$ 
in the FF of Hamiltonian dynamics is known to be 
related to the vacuum issue in QFT. In the context of 
renormalization group procedure in QCD, the issue is 
discussed in Ref.~\cite{Wilsonetal}. But it has a long 
history~\cite{Casher:1974xd}, including the possibility 
that the so-called vacuum condensates are reducible 
to universal small-$x$ components of hadrons. This 
possibility leads to the idea that the so-called vacuum 
condensates are associated with the hadron states 
rather than the vacuum state~\cite{Maris:1997hd,
Brodsky:2010xf,GlazekCondensatesAPP,Brodsky:2012ku}. 
A free massless particle has the FF ``energy'' $k^- 
= k^{\perp \,2}/k^+$, where $k^\pm = k^0 \pm k^3$. 
The limit of small $k^\perp$ and $ x = k^+/ p^+$, with 
$p^+$ denoting the hadron momentum, does not imply 
any definite value of $k^-$ and all scales of the ``energy'' 
$k^-$ are mixed up in the IR dynamics. Once a mass scale 
$\kappa$ is introduced,  the FF ``energy'' becomes $k^- 
= (k^{\perp \,2}+ \kappa^2)/k^+$ and the small parton 
fraction $x$ for fixed hadron $p^+$ necessarily implies a 
large $k^-$. Since it is known that renormalization of 
singularities at small $k^+$, related to small $x$, may 
shed new light on the otherwise perplexing vacuum 
issue~\cite{Wilsonetal}, even the slightest possibility of 
introducing the mass scale $\kappa$ as a regulating 
parameter for gauge bosons, resolving the intricate FF 
scale mixing and turning the limit of small $x$ into 
an UV one, deserves to be noted. Serving this purpose,
this article is of technical nature. 

Section~\ref{LD} introduces the theory we consider. 
In addition to fermions and gauge bosons, a complex 
scalar field is introduced. Its potential has a minimum 
at a definite value of the field modulus, while the field 
phase is arbitrary. A limiting theory, for brevity called 
{\it massive}, is identified. Later on, this theory is used 
to derive the quantum Hamiltonian for massive gauge 
bosons interacting with fermions. In Sec.~\ref{TCG}, 
following Soper's work~\cite{Soper}, two different choices 
of gauge are described. One choice shows that the 
Lagrangian density we consider corresponds to a theory 
of massive vector bosons coupled to fermions. Another 
one is suitable for constructing a FF quantum Hamiltonian 
for vector bosons coupled to fermions. Sec.~\ref{TCG} 
provides details of the two gauge choices. The FF Hamiltonian 
density in gauge $A^+=0$ is calculated in Sec.~\ref{CofH}, 
prior to taking the massive limit. The calculation involves 
solving constraint equations for fermion and gauge-boson 
fields. The solution yields dynamics that involves the 
modulus field. Sec.~\ref{HML} employs the massive 
limit to derive the Hamiltonian for fermions coupled to 
gauge bosons only, while the modulus field is decoupled. 
The corresponding quantum Hamiltonian is obtained using 
standard FF field quantization procedure. Sec.~\ref{polarization}
explains how the phase field provides the third polarization 
state for gauge vector bosons and Sec.~\ref{C} concludes 
the paper with comments concerning regularization.
 
\section{ Lagrangian density }
\label{LD}

The gauge theory we discuss is introduced in terms of a  
local Lagrangian density of the familiar structure 
\beq
\label{cL}
\cL \es \cL_\psi + \cL_A + \cL_{A\phi} - \cV_{\phi} \ , 
\eeq 
where 
\beq
\label{Lpsi}
\cL_\psi 
\es \bar \psi \left[ \left( i \partial_\mu - g A_\mu \right)
                            \gamma^\mu - m \right] \psi \ , \\
\label{LA}
\cL_A 
\es - {1 \over 4} F_{\mu \nu} F^{\mu \nu} \ , \\
\label{Aphi}
\cL_{A \phi} 
\es \left[ (i\partial^\mu - g' A^\mu ) \phi \right]^\dagger 
                            (i\partial_\mu - g' A_\mu ) \phi  \ , \\
\label{Lphi}
\cV_\phi
\es
-\mu^2 \ \phi^\dagger \phi + {\lambda^2 \over 2} \ ( \phi^\dagger \phi )^2  \ .
\eeq
It involves a fermion field $\psi$, an Abelian vector field $A$ with
field-strength tensor $F_{\mu \nu} = \partial_\mu A_\nu - \partial_\nu A_\mu$,
and a complex scalar field $\phi$ with a real potential $\cV_\phi$
in which the constant $\mu^2$ is meant to be positive.
The Lagrangian density exhibits gauge symmetry in the 
sense that it does not change its form when one makes 
the replacements,
\beq
\psi       \tl  e^{ig f} \psi  \ , \\
A^\mu  \tl  A^\mu - \partial^\mu f \ , \\
\phi      \tl  e^{ig'f} \phi \ ,
\eeq
where $f$ is a function of points in space-time. 

The complex field $\phi$ does not interact directly 
with the fermion field, but it does interact with fermions
indirectly, through the gauge field $A$. When the interaction
of field $\phi$ with gauge field $A$ is turned off, {\it i.e.}, 
when $g'$ is set to zero, the field $\phi$ becomes a real 
self-interacting field that does not interact with any other 
field. The Yukawa-like coupling $\bar \psi \phi \psi$ is
excluded by the requirement of gauge invariance. 

\subsection{ Phase field $\theta$ }

The field $\phi$ can be written using its modulus $|\phi| = 
\varphi/ \sqrt{2}$ and phase $g' \theta$~\cite{Kibble},
\beq
\phi  \es  \varphi \ e^{i g' \theta} /\sqrt{2} \ .
\eeq
The real field $\varphi$ can be considered constant and denoted 
as such by $v$, as if it represented a ground-state, or vacuum 
expectation value of a field operator in a quantum theory.  In the 
theory we consider, the constant $v$ is just a free parameter. The 
field $\varphi$ can deviate from the constant $v$. Such deviation 
is denoted here by $h$, in analogy to the higgs field in the standard 
model  (SM). The potential $\cV(\phi)$ is a function of $\varphi$ 
alone, $\cV(\phi) = \cV(\varphi/ \sqrt{2})$; it does not depend on 
the phase field $\theta$. 

When $\varphi = v + h$ and $h=0$, the 
potential has its minimal value $-\mu^4/(2\lambda^2)$ 
for $v = \sqrt{2} \ \mu/ \lambda$. Using this special value 
of $v$ for $h \neq 0$, one has
\beq
\cV(\phi)
\rs
- { \mu^4 \over 2 \lambda^2} + \2 \, (\sqrt{2} \, \mu )^2 \ h^2 
+ { \lambda \over \sqrt{2} } \ \mu \ h^3 + {\lambda^2 \over 8} \ h^4  \ ,
\eeq
which means that a term linear in $h$ is absent. 
The Lagrangian density depends on the gradient 
$\partial^\mu \theta$;
\beq
\label{Aphi2}
\cL_{A \phi} 
\es \2 (\partial^\mu \varphi )^2 
     +
     \2 
     g'^2 (A^\mu + \partial^\mu \theta )^2  \varphi^2 \ .
\eeq

\subsection{ Massive limit }
\label{ML}

We will consider the gauge theory in the limit of $g' \to 0$, 
$v \to \infty$ and $g' v = \kappa$ kept constant. To be short,
we call this limit the {\it massive} limit. For the sake of further
discussion, the massive limit is specified by assuming that the 
positive mass parameter $\mu$ is fixed, but arbitrary, and 
the field $\varphi$ self-interaction coupling constant $\lambda 
\to 0$. In this limit the field $h$ can be neglected in comparison 
to the constant $v$ for as long as the divergences caused by the 
field $h$ are regulated and negligible in comparison with the
corresponding powers of the constant $v$. Hence, $v + h 
\to v$, $g'\phi \to \kappa$ and 
\beq
\label{Aphimassive}
\cL_{A \phi} 
\es \2 ( \partial^\mu h )^2 
     +
     \2 
     \kappa^2 (A^\mu + \partial^\mu \theta )^2 \ , \\
\label{Lphi2massive}
\cV_\phi
\es
- { \mu^4 \over 2 \lambda^2} 
+ \2 \, (\sqrt{2} \, \mu )^2 \ h^2 
\ .
\eeq
The field $h$ becomes free, with mass $\sqrt{2} \ \mu$, and 
the field $A^\mu + \partial^\mu \theta$ obtains the mass $\kappa$.  
Since the Yukawa-like coupling $\bar \psi \phi \psi$ is
excluded, the modulus field cannot generate the mass
of fermions.

\subsection{ Gauge symmetry in terms of $\psi$, $A^\mu$, $\varphi$ and $\theta$ }

In summary, the Lagrangian density written in terms of fields 
$\psi$, $A^\mu$, $\varphi$ and $\partial^\mu \theta$ is 
$ \cL = \cL_\psi + \cL_A + \cL_{A\phi} - \cV_{\phi}$,
where
\beq
\label{Lpsi3}
\cL_\psi 
\es \bar \psi \left[ \left( i \partial_\mu - g A_\mu \right)
                            \gamma^\mu - m \right] \psi \ , \\
\label{LA3}
\cL_A 
\es - {1 \over 4} F_{\mu \nu} F^{\mu \nu} \ , \\
\label{Aphi3}
\cL_{A \phi} 
\es \2 (\partial^\mu \varphi )^2 
     +
     \2 
     g'^2 (A^\mu + \partial^\mu \theta )^2  \varphi^2  \ , \\
\label{Lphi3}
\cV_\phi
\es
\cV(\varphi/\sqrt{2}) \ .
\eeq
The gauge transformation that leaves the Lagrangian density
unchanged is
\beq
\psi       \tl  e^{ig f} \psi  \ , \\
A^\mu  \tl  A^\mu - \partial^\mu f \ , \\
\varphi  \tl \varphi \ , \\
\theta   \tl \theta + f  \ .
\eeq
The transformation is realized by substitutions
\beq
\label{gaugepsi}
\psi       \es   e^{-ig f} \tilde \psi  \ , \\
\label{gaugeA}
A^\mu   \es  \tilde A^\mu + \partial^\mu f \ , \\
\label{gaugevarphi}
\varphi  \es \tilde \varphi \ , \\
\label{gaugetheta}
\theta   \es \tilde \theta  - f  \ .
\eeq
The Lagrangian density as a function of fields without
tilde and as a function of fields with a tilde is the same
function.  Two different choices for the function $f$
are used in what follows.

\section{ Two choices of gauge }
\label{TCG}

Following Soper~\cite{Soper}, we use two different 
choices of gauge in order to obtain two results. In the
first gauge, the result is that the Lagrangian density we 
consider corresponds to a theory of massive vector 
fields coupled to fermions. In the second gauge, an 
explicit derivation of quantum Hamiltonian for the 
theory becomes possible.  The difference from~\cite{Soper}
is that the mass parameter, which Soper introduces in 
order to obtain gauge symmetry and which he denotes 
by $\kappa$, is in our gauge theory example obtained 
as a result of a different gauge symmetry, which includes 
an additional field $h$ in a way resembling but not identical 
to the Higgs mechanism in the SM. The field $h$ is not 
involved in generating fermion masses. Our parameter 
$\kappa$ that corresponds to Soper's, appears from 
substitution $\kappa = g'v$. The additional field $h$ 
decouples only in the massive limit.

\subsection{ Gauge choice $f = - \theta$ }
\label{cLAtheta}

For $f = - \theta$, we have 
\beq
\psi       \tl  \tilde \psi  \rs e^{ig f} \psi  \rs e^{-ig \theta} \psi \ , \\
A^\mu  \tl  \tilde A     \rs A^\mu - \partial^\mu f  \rs A^\mu + \partial^\mu \theta \ , \\
\varphi  \tl  \tilde \varphi \rs \varphi \ , \\
\theta   \tl  \tilde \theta \rs  \theta + f \rs 0  \ .
\eeq
These substitutions are realized by setting
\beq
\psi       \es  e^{ig \theta} \tilde \psi \ , \\
A^\mu   \es  \tilde A - \partial^\mu \theta \ , \\
\varphi   \es  \tilde \varphi \ , \\
\theta    \es  \tilde \theta + \theta  \ ,
\eeq
which leads to $ \cL = \cL_\psi + \cL_A + \cL_{A\phi} 
- \cV_{\phi}$ with
\beq
\label{Lpsitheta1}
\cL_\psi 
\es \bar {\tilde \psi} \left[ \left( i \partial_\mu - g \tilde A_\mu \right)
                            \gamma^\mu - m \right] \tilde \psi \ , \\
\label{LA3theta1}
\cL_A 
\es - {1 \over 4} \tilde F_{\mu \nu} \tilde F^{\mu \nu} \ , \\
\label{Aphi3theta1}
\cL_{A \phi} 
\es 
     \2 (\partial^\mu \tilde \varphi )^2 
     +
     \2 
     g'^2 \tilde A^{\mu \, 2}  \tilde \varphi^2
     \ , \\
\label{Lphi3theta1}
\cV_\phi
\es
\cV ( \tilde \varphi/\sqrt{2}) \ .
\eeq
In the massive limit, $\cL_\psi$, $\cL_A$
remain the same, $\cL_{A\phi}$ becomes
\beq
\cL_{A \phi} 
\es \2 (\partial^\mu \tilde h)^2 
     +
     \2 
     \kappa^2 \tilde A^2   \ ,
\eeq
and $\cV_\phi$ is dominated by the constant $\mu^2/(2 \lambda^2)$.
Thus, in the massive limit the Lagrangian density is
$ \cL = \cL_\psi + \cL_A + \cL_{A\phi} - \cV_{\phi}$ , where
\beq
\label{Lpsitheta2}
\cL_\psi 
\es \bar {\tilde \psi} \left[ \left( i \partial_\mu - g \tilde A_\mu \right)
                            \gamma^\mu - m \right] \tilde \psi \ , \\
\label{LA3theta2}
\cL_A 
\es - {1 \over 4} \tilde F_{\mu \nu} \tilde F^{\mu \nu} \ , \\
\label{Aphi3theta2}
\cL_{A \phi} 
\es \2 (\partial^\mu \tilde h)^2 
     +
     \2 
     \kappa^2 \tilde A^2
     \ , \\
\label{Lphi3theta2}
\cV_\phi
\es
- { \mu^2 \over 2 \lambda^2 } 
+ \2 \, (\sqrt{2} \, \mu )^2 \ \tilde h^2 \ .
\eeq
This Lagrangian density yields the action of a free field $\tilde h$ of
mass $\sqrt{2} \, \mu$ and, additively, of a massive vector field $\tilde A$ 
coupled to the fermion field $\tilde \psi$. Assuming that the gauge 
symmetry is realized in nature and that photons coupled to charged 
fermions have a very small mass, corresponding to a very small $\kappa$, 
there also ought to exist in nature a scalar field $h$ with an unknown 
mass whose value of $\sqrt{2} \, \mu$ is not in any way limited by the 
theory. A different form of the potential density $\cV_\phi$ than in 
Eq.~(\ref{Lphi}), would lead to a different mass of the field $h$ in the 
massive limit that is realized through some analog of $\lambda \to 0$.

Current upper limit on the photon mass~\cite{PDG} is $10^{-18}$ 
eV/$c^2$, which is extremely low. There are no data suggesting that 
the field $h$ exists, but the astrophysical data stimulate searches for 
the dark matter and other exotic particles~\cite{PDG}.

\subsection{ Gauge choice $\tilde A^+=0$ }
\label{cLA+}

The original Lagrangian density allows for simultaneous 
alteration of the fermion field $\psi$ and phase field 
$\theta$ so that the vector field $A$ can be replaced 
by a similar field $\tilde A$ whose component $\tilde 
A^+ = 0$. The $+$-component of a four-vector is defined 
in the same way as for all tensors, $\pm = 0 \pm 3$. 
For example, a position four-vector $x$ has components 
$(x^-,x^+,x^1,x^2)$ and the gradient components are 
$(\partial^-,\partial^+,\partial^\perp)$. Components 1 
and 2 are collectively denoted by $\perp$. 

The choice of $\tilde A^+=0$ stems from the form of 
dynamics that we use to construct the Hamiltonian~\cite{DiracFF}. 
We use the FF, instead of the commonly used form that 
Dirac called the instant form (IF)~\cite{DiracFF}. Setting 
$\tilde A^+=0$ is useful because it leads to simple and 
soluble constraint equations. Also, it is invariant with respect
to seven kinematic Poincar\'e transformations that form 
the FF symmetry group  in the Minkowsky space-time.
The IF has only six kinematic symmetries~\cite{FFreview}.

Since the couplings of $A$ to $\psi$ and $\phi$ 
have different strengths, given by different dimensionless
coupling constants $g$ and $g'$, different changes of phase 
are needed for fields $\psi$ and $\phi$ to obtain $\tilde A^+ 
= 0$. One can introduce the function $f$ that satisfies the 
equation
\beq
\partial^+ f \es A^+ \ .
\eeq
For example, 
\beq
f(x) \es \, {1 \over 4}  \left(  \int_{-\infty}^{x^-} - \int_{x^-}^\infty \right) dy^- A^+(x^+,y^-,x^\perp) \ .
\eeq
An arbitrary function that does not depend on $x^-$ could be added
to this definition. The gradient of this $f$ is
\beq
\partial^- f (x)\es  {1 \over 4}  \left(  \int_{-\infty}^{x^-} - \int_{x^-}^\infty \right) dy^- 
\partial^- A^+(x^+,y^-,x^\perp) \ , \\
\partial^+ f (x)\es A^+(x^+,x^-,x^\perp) \ , \\
\partial^\perp f (x)\es {1 \over 4}  \left(  \int_{-\infty}^{x^-} - \int_{x^-}^\infty \right) dy^- 
\partial^\perp A^+(x^+,y^-,x^\perp) \ .
\eeq
The gauge transformation is realized by substitutions
\beq
\psi       \es  e^{-ig f} \tilde \psi \ , \\
A^\mu  \es  \tilde A^\mu + \partial^\mu f \ , \\
\varphi  \es  \tilde \varphi \ , \\
\theta   \es  \tilde \theta -  f  \ ,
\eeq
which produce the same Lagrangian density 
$ \cL = \cL_\psi + \cL_A + \cL_{A\phi} - \cV_{\phi}$
as the one in Eqs.~(\ref{Lpsi3}), (\ref{LA3}), 
(\ref{Aphi3}) and (\ref{Lphi3}) except that the fields
$\psi$, $A$, $\varphi$ and $\theta$ are replaced
by fields  $\tilde \psi$, $\tilde A$, $\tilde \varphi$ 
and $\tilde \theta$,  {\it i.e.}, 
\beq
\label{Lpsi3+}
\cL_\psi 
\es \bar {\tilde \psi} \left[ \left( i \partial_\mu - g \tilde A_\mu \right)
                            \gamma^\mu - m \right] \tilde \psi \ , \\
\label{LA3+}
\cL_A 
\es - {1 \over 4} \tilde F_{\mu \nu} \tilde F^{\mu \nu} \ , \\
\label{Aphi3+}
\cL_{A \phi} 
\es \2 (\partial^\mu \tilde \varphi )^2 
     +
     \2 
     g'^2 (\tilde A^\mu + \partial^\mu \tilde \theta )^2  \tilde \varphi^2  \ , \\
\label{Lphi3+}
\cV_\phi
\es
\cV[\tilde \varphi/\sqrt{2}] \ ,
\eeq
and $\tilde A^+=0$. In the massive limit,
\beq
\label{Lpsi3+m}
\cL_\psi 
\es \bar {\tilde \psi} \left[ \left( i \partial_\mu - g \tilde A_\mu \right)
                            \gamma^\mu - m \right] \tilde \psi \ , \\
\label{LA3+m}
\cL_A 
\es - {1 \over 4} \tilde F_{\mu \nu} \tilde F^{\mu \nu} \ , \\
\label{Aphi3+m}
\cL_{A \phi} 
\es \2 (\partial^\mu \tilde h )^2 
     +
     \2 
     \kappa^2 (\tilde A^\mu + \partial^\mu \tilde \theta )^2  \ , \\
\label{Lphi3+m}
\cV_\phi
\es
- { \mu^2 \over 2 \lambda^2 } 
+ \2 \, (\sqrt{2} \, \mu )^2 \ \tilde h^2 \ ,
\eeq
which look the same as Eqs.~(\ref{Lpsitheta2}), 
(\ref{LA3theta2}), (\ref{Aphi3theta2}) and (\ref{Lphi3theta2}).
However, the component $\tilde A^+$ of the gauge field is 
zero. Instead, the gradient of field $\tilde \theta$ is present. 

\section{ Calculation of the FF Hamiltonian }
\label{CofH}

The Lagrangian density $ \cL = \cL_\psi + \cL_A + 
\cL_{A\phi} - \cV_{\phi}$ of Eqs.~(\ref{Lpsi}),
(\ref{LA}), (\ref{Aphi}) and (\ref{Lphi}) is written
in Secs.~\ref{cLAtheta} and \ref{cLA+} in two 
formally equivalent versions that differ by the 
choice of gauge. In the massive limit, the version
of Sec.~\ref{cLAtheta} coincides with a theory of 
a fermion field coupled in a mininmal way to a 
massive vector field, while the version of Sec.~\ref{cLA+} 
coincides with a theory of a fermion field coupled
to a massive transverse vector field and this vector 
field is coupled to a gradient of a massive scalar field. 

These two formally equivalent massive limit versions 
of the gauge theory correspond to the theory developed 
by Soper~\cite{Soper}. In this section a FF Hamiltonian 
that corresponds to the Lagrangian $\cL$ is constructed 
using the gauge $\tilde A^+=0$ of Sec.~\ref{cLA+} 
before the massive limit is taken. The resulting FF 
Hamiltonian formally reduces in the massive limit to 
the Soper FF Hamiltonian .

\subsection{ Equations of motion }

The Lagrangian density of Eqs.~(\ref{Lpsi}) to (\ref{Lphi})
implies, through the principle of minimal action, the 
Euler-Lagrange (EL) equations of motion,
\beq
\label{ELpsi}
\left[ \left( i \partial_\mu - g A_\mu \right)
                            \gamma^\mu - m \right] \psi \es 0 \ , \\
\label{ELA}
 - \partial_\alpha ( \partial^\alpha A^\beta - \partial^\beta A^\alpha ) 
\es 
- g \bar \psi \gamma^\beta \psi 
+
g'^2 \ \varphi^2  \ ( A^\beta + \partial^\beta \theta  ) \ , \\
\label{ELvarphi}
\partial_\mu \partial^\mu \varphi 
\es
g'^2 \ \varphi  \ ( A^\beta + \partial^\beta \theta  )^2 
- {\partial \cV(\varphi/\sqrt{2}) \over \partial \varphi} \ , \\
\label{ELtheta}
\partial_\mu
 g'^2 (A^\mu + \partial^\mu \theta )  \varphi^2 
\es 0 \ .
\eeq
The last equation is necessarily satisfied if the first 
two are. In the massive limit, these equations become
\beq
\label{mELpsi}
\left[ \left( i \partial_\mu - g A_\mu \right)
                            \gamma^\mu - m \right] \psi \es 0 \ , \\
\label{mELA}
 - \partial_\alpha ( \partial^\alpha A^\beta - \partial^\beta A^\alpha ) 
\es 
- g \bar \psi \gamma^\beta \psi 
+
\kappa^2  \ ( A^\beta + \partial^\beta \theta  ) \ , \\
\label{mELvarphi}
\partial_\mu \partial^\mu h
\es
- {\partial \cV(h/\sqrt{2}) \over \partial h } \ , \\
\label{mELtheta}
\kappa^2 \
\partial_\mu (A^\mu + \partial^\mu \theta ) 
\es 0 \ .
\eeq
Assuming non-zero mass $\kappa$ and introducing the field
\beq
B \es - \kappa \ \theta \ ,
\eeq
one obtains the massive EL equations in the form
\beq
\label{mELpsi1}
\left[ \left( i \partial_\mu - g A_\mu \right)
                            \gamma^\mu - m \right] \psi \es 0 \ , \\
\label{mELA1}
\Box A^\beta - \partial^\beta \partial_\alpha A^\alpha 
\es 
g \bar \psi \gamma^\beta \psi 
-
\kappa^2  \ ( A^\beta  - \kappa^{-1} \partial^\beta B  ) \ , \\
\label{mELtheta1}
\kappa^2 \
\partial_\mu (A^\mu - \kappa^{-1} \partial^\mu B ) 
\es 0 \ , \\
\label{mELvarphi1}
\partial_\mu \partial^\mu h
\es
- {\partial \cV(h/\sqrt{2}) \over \partial h } \ .
\eeq
Note that $\Box B = \kappa \partial_\mu A^\mu$ 
and the field $h$ is decoupled. Equations (\ref{mELpsi1}),
(\ref{mELA1}) and (\ref{mELtheta1}) for fields $\psi$, 
$A$ and $B$ coincide with Eqs. (2), (3) and (4) of
 Soper~\cite{Soper}. Equation (\ref{mELvarphi1}) is absent
 in Soper's theory. The massive limits described in 
Secs.~\ref{cLAtheta} and \ref{cLA+} lead to the EL
equations that coincide with Soper's equations in 
gauges $B=0$ and $A^+=0$, respectively. Our 
construction of the FF Hamiltonian for massive gauge 
bosons is carried out in the gauge $\tilde A^+=0$. 

Prior to taking the massive limit, the EL equations 
in terms of the field $B$ read
\beq
\label{ELpsiB}
\left[ \left( i \partial_\mu - g A_\mu \right)
                            \gamma^\mu - m \right] \psi \es 0 \ , \\
\label{ELAB}
\Box A^\beta - \partial^\beta \,  \partial_\alpha A^\alpha  
\es 
g \bar \psi \gamma^\beta \psi 
-
g'^2 \ \varphi^2  \ ( A^\beta - \kappa^{-1} \partial^\beta B  ) \ , \\
\label{ELvarphiB}
\Box \varphi 
\es
g'^2 \ \varphi  \ ( A^\beta - \kappa^{-1} \partial^\beta B )^2 
- {\partial \cV(\varphi/\sqrt{2}) \over \partial \varphi} \ , \\
\label{ELthetaB}
\partial_\mu
 g'^2 \varphi^2  (A^\mu - \kappa^{-1}  \partial^\mu B)  
\es 0 \ .
\eeq
Again, the last equation must be satisfied if the first two are.

\subsection{ Constraint equations in gauge $\tilde A^+=0$ }

In the FF of dynamics, the constraint equations are those
EL equations that do not involve differentiation with respect
to $x^+$, {\it i.e.}, $\partial^- = 2 \partial/\partial x^+$. We 
abbreviate our notation from $\tilde \psi$, $\tilde A$, 
$\tilde \varphi$ and $\tilde \theta$ to $\psi$, $A$, 
$\varphi$ and $\theta$.

\subsubsection{ Constraint equations for the fermion field }

The $4 \times 4$ projection matrices $\Lambda_\pm = 
\2 \gamma^0 \gamma^\pm = \2 ( 1 \pm \alpha^3)$, 
which have the properties $\Lambda_\pm \alpha^\perp 
= \alpha^\perp \Lambda_\mp$ and $\Lambda_\pm 
\beta = \beta \Lambda_\mp$, are used to write the 
fermion field as $\psi = \psi_+ + \psi_-$, where $\psi_\pm 
= \Lambda_\pm \psi$. The fermion EL Eq.~(\ref{ELpsi}) 
takes the form 
\beq
 \left[   (i \partial^- -g A^-) \Lambda_+ +  i \partial^+ \Lambda_- 
                          - ( i\partial^\perp-gA^\perp)  \alpha^\perp - m\beta \right] \psi \es 0 \ ,
\eeq
which consists of two coupled equations  for $\psi_+$ and $\psi_-$,
\beq
\label{psi1}
(i \partial^- -g A^-) \psi_+  
                          - \left[ ( i\partial^\perp-gA^\perp)  \alpha^\perp + m\beta \right] \psi_- \es 0 \ , \\
\label{psi2}
i \partial^+ \psi_- 
                          - \left[ ( i\partial^\perp-gA^\perp)  \alpha^\perp + m\beta \right] \psi_+ \es 0 \ .                          
\eeq
The second equation yields
\beq
\label{psiminus}
 \psi_- \es {1 \over i \partial^+} \
             \left[ ( i\partial^\perp-gA^\perp)  \alpha^\perp + m\beta \right] \psi_+ \ ,                   
\eeq
so that the first one describes the evolution of $\psi_+$ in $x^+$,
\beq
\label{psievo}
i \partial^- \psi_+
\es
\left\{ 
\left[ ( i\partial^\perp-gA^\perp)  \alpha^\perp + m\beta \right] 
\right.
\nt  
\left.
{1 \over i \partial^+} 
\left[ ( i\partial^\perp-gA^\perp)  \alpha^\perp + m\beta \right] 
+
g A^- \right\}
\psi_+
\ ,
\eeq
while $\psi_-$ is given by the constraint Eq.~(\ref{psiminus}).

\subsubsection{ Constraint equation for the boson field }

The EL Eq.~(\ref{ELA}) for the field $A$ is
\beq
\Box A^\beta - \partial^\beta \partial_\alpha A^\alpha 
\es 
g \bar \psi \gamma^\beta \psi 
-
g'^2 \ \varphi^2  \ ( A^\beta + \partial^\beta \theta  ) \ ,
\eeq
which in terms of the field $B = -\kappa \theta$ reads
\beq
\Box A^\beta - \partial^\beta \partial_\alpha A^\alpha 
\es 
g \bar \psi \gamma^\beta \psi 
-
g'^2 \ \varphi^2  \ ( A^\beta - \kappa^{-1} \partial^\beta B  ) \ .
\eeq
Setting $\beta = +$, one gets in gauge $A^+=0$ that
\beq
\partial^+ \partial_\alpha A^\alpha 
\es 
- g \bar \psi \gamma^+ \psi 
-
g'^2 \ \varphi^2 \  \kappa^{-1} \partial^+ B  
\ .
\eeq
This equation contains no derivatives with respect to $x^+$
and it is a constraint. One gets
\beq
 \partial_\alpha A^\alpha 
\es - \, {1 \over \partial^+ }
\left( g \bar \psi \gamma^+ \psi 
    + g'^2 \ \varphi^2 \ \kappa^{-1} \partial^+ B  
\right)
\ ,
\eeq
and, as a result of $ \partial_\alpha A^\alpha = \partial^+ A^-/2 - \partial^\perp A^\perp $,
\beq
\label{constraintA}
A^- \es { 2 \over \partial^+ } \partial^\perp A^\perp
-
{2 \over \partial^{+ \, 2} }
\left( g \bar \psi \gamma^+ \psi 
+
g'^2 \ \varphi^2 \  \kappa^{-1} \partial^+ B  
\right) \ .
\eeq
This constraint makes the evolution of $\psi_+$ in Eq.~(\ref{psievo}) nonlinear.
It also shows that the $(1/\partial^+)B$ couples to fermions as $A^-$ does.

\subsection{ Lagrangian density in gauge $A^+=0$ }

By inspection, one finds that the Lagrangian density $\cL$
in the gauge $A^+=0$ is linear in derivatives of fields with
respect to $x^+$. The fermion Lagrange density is
\beq
\cL_\psi
\es
\psi_+^\dagger (i \partial^- -g A^-)\psi_+
+
\psi_-^\dagger i \partial^+ \psi_-
-
\psi_+^\dagger \left[ ( i\partial^\perp-gA^\perp)  \alpha^\perp + m\beta \right] \psi_- 
\nm
\psi_-^\dagger \left[ ( i\partial^\perp-gA^\perp)  \alpha^\perp + m\beta \right] \psi_+
\ .
\eeq
The gauge field density is
\beq
{- 1 \over 4} F_{\mu \nu} F^{\mu \nu} 
\es 
{- 1 \over 4}
\left[
- \, {1 \over 2} (\partial^+ A^-)^2
+ 2  (\partial^k A^- - \partial^- A^k) \partial^+ A^k
\right.
\np
\left.
 2 \partial^l A^k \partial^l A^k - 2 \partial^l A^k \partial^k A^l 
\right] 
\ .
\eeq 
The densities for fields $\varphi$ and $\partial^\mu B$ are
\beq
\cL_{A \phi} 
\es \2 \left( \partial^+ \varphi \ \partial^- \varphi  - \partial^\perp \varphi \ \partial^\perp \varphi \right)
     \np
     \2 
     g'^2   \left[ - \kappa^{-1} \partial^+ B (A^- - \kappa^{-1} \partial^- B )   
     - (A^\perp - \kappa^{-1} \partial^\perp B )^2
      \right] \varphi^2  \ , \\
\cV_\phi
\es
\cV(\varphi/\sqrt{2}) \ .
\eeq

\subsection{ Derivation of the FF Hamiltonian density }

The energy-momentum tensor, generally given by
\beq
T^{\mu \nu} 
\es
\sum_i 
       { \partial \cL \over \partial \partial_\mu f_i } \
       \partial^\nu f_i 
-
g^{\mu \nu}  \cL  \ ,
\eeq
leads to the Hamiltonian $P^-$,
\beq
P^- \es {1 \over 2} \int d^2 x^\perp dx^- \  T^{+ \, -} \ .
\eeq
Therefore, 
\beq
\label{Pminus}
P^- \es \int d^2 x^\perp dx^- \ 
\left(
\sum_i 
       { \partial \cL \over \partial \partial^- f_i } \
       \partial^- f_i  - \cL 
\right) \ ,
\eeq
For Lagrangian densities that are linear in partial derivatives
$\partial^-$ of fields, the Hamiltonian densities $\cH$ are 
simply given by the formula $\cH = - \cL(\partial^- \to 0)$. 
This simplification occurs for all Lagrangian densities that 
are quadratic in the IF time derivatives and obey the principles 
of special relativity. In our case,
\beq
\cH \es 
g \psi_+^\dagger A^-\psi_+
-
\psi_-^\dagger i \partial^+ \psi_-
\np
\psi_+^\dagger \left[ ( i\partial^\perp-gA^\perp)  \alpha^\perp + m\beta \right] \psi_- 
+
\psi_-^\dagger \left[ ( i\partial^\perp-gA^\perp)  \alpha^\perp + m\beta \right] \psi_+
\np
{1 \over 4} 
\left[
- \, {1 \over 2} (\partial^+ A^-)^2
+ 2  \partial^k A^- \partial^+ A^k
+ 2 \partial^l A^k \partial^l A^k - 2 \partial^l A^k \partial^k A^l
\right]
\np \2 (\partial^\perp \varphi)^2
+ \2 g'^2 \varphi^2 
\left[ \kappa^{-1} \partial^+ B \ A^-
+ (A^\perp - \kappa^{-1} \partial^\perp B )^2 \right]
+ \cV (\varphi/\sqrt{2}) \ , \nn
\eeq
where for the optimal choice of $v$ in $\varphi = v + h$,
\beq
\cV(\varphi/\sqrt{2}) 
\es
- { \mu^4 \over 2 \lambda^2} 
+ \2 \, (\sqrt{2} \, \mu )^2 \ h^2 
+ h^2 \ \left[ { \lambda \over \sqrt{2} } \ \mu \ h +  {\lambda^2 \over 8} \ h^2 \right] \ .
\eeq
Constraints to include are
\beq
 \psi_- \es {1 \over i \partial^+} \
               \left[ ( i\partial^\perp-gA^\perp)  \alpha^\perp + m\beta \right] \psi_+ \ , \\
A^- \es { 2 \over \partial^+ } \partial^\perp A^\perp
-
{2 \over \partial^{+ \, 2} }
\left[ 
g \bar \psi \gamma^+ \psi 
+
g'^2 \ \varphi^2 \ \kappa^{-1}  \partial^+ B  
\right]  \ .      
\eeq
The field $\psi_-$ is most easy to eliminate.
Looking at the constraint Eq.~(\ref{psiminus}), 
one sees that the terms 
$-\psi_-^\dagger i \partial^+ \psi_-$ and 
$\psi_-^\dagger \left[ ( i\partial^\perp-gA^\perp)  \alpha^\perp + m\beta \right] \psi_+ $
cancel each other and one is left with
\beq
\psi_+^\dagger \left[ ( i\partial^\perp-gA^\perp)  \alpha^\perp - m\beta \right]  
{1 \over i \partial^+} \
\left[ ( i\partial^\perp-gA^\perp)  \alpha^\perp + m\beta \right] \psi_+ \ .
\eeq
Turning to the vector field, the four terms that involve $A^-$,
\beq
\2 g \bar \psi \gamma^+ \psi A^-
+
{1 \over 4} 
\left[
- \, {1 \over 2} (\partial^+ A^-)^2
+ 2  \partial^k A^- \partial^+ A^k \right]
+ 
\2 g'^2 \varphi^2 \ A^-\ \kappa^{-1}\partial^+B  \ ,  \nn
\eeq
can be shown, using partial integration over the front,  
to be equivalent to 
\beq
{1 \over 8} ( \partial^+ A^- )^2 \ .
\eeq
The full Hamiltonian is then
\beq
\label{cH1}
\cH \es 
\2 ( \partial^+ A^-/2 )^2
\np
\psi_+^\dagger \left[ ( i\partial^\perp-gA^\perp)  \alpha^\perp + m\beta \right]  
{1 \over i \partial^+} \
\left[ ( i\partial^\perp-gA^\perp)  \alpha^\perp + m\beta \right] \psi_+
\np
\2 
\left( \partial^l A^k \partial^l A^k - \partial^l A^k \partial^k A^l \right)
\np \2 (\partial^\perp \varphi )^2
+ \2 g'^2 \varphi^2 (A^\perp - \kappa^{-1} \partial^\perp  B )^2 
+ \cV \left( \varphi/\sqrt{2} \right) \ ,
\eeq
where
\beq
A^- \es {2 \over \partial^+} \partial^\perp A^\perp 
-
{2 \over \partial^{+ \, 2} }
\left[ 
g \bar \psi \gamma^+ \psi 
+
g'^2 \varphi^2 \kappa^{-1}  \partial^+ B
\right]  \ .
\eeq
The FF Hamiltonian density of Eq.~(\ref{cH1}) is 
used below to derive the Hamiltonian for massive 
gauge bosons coupled to fermions, taking 
advantage of  the massive limit.

\section{  Hamiltonian in the massive limit }
\label{HML}

In the massive limit, which is defined in Sec.~\ref{ML},
the coupling constant $g' \to 0$ and the modulus 
parameter $v \to \infty$ with the product $g'v = 
\kappa$ kept constant. One has 
\beq
g' \varphi \es g' v(1 + h/v) \rs \kappa ( 1 + h/v) \to \kappa \ .
\eeq
Recall that in terms of the Lagrangian density of Eq.~(\ref{Lphi}),
the massive limit is set by demanding that the coupling constant 
$\lambda \to 0$ for an arbitrary but fixed value of the mass 
parameter $\mu$. Thus, in the massive limit defined by both 
$g'$ and $\lambda$ tending to zero, the Hamiltonian density 
of Eq.~(\ref{cH1}) becomes
\beq
\label{cH2}
\cH \es
\2 ( \partial^+ A^-/2 )^2
\np
\psi_+^\dagger \left[ ( i\partial^\perp-gA^\perp)  \alpha^\perp + m\beta \right]  
{1 \over i \partial^+} \
\left[ ( i\partial^\perp-gA^\perp)  \alpha^\perp + m\beta \right] \psi_+
\np
\2 
\left( \partial^l A^k \partial^l A^k - \partial^l A^k \partial^k A^l \right)
\np 
\2 \kappa^2 (A^\perp - \kappa^{-1} \partial^\perp  B )^2  
+ \2 \, h  ( - \partial^{\perp \, 2} + 2 \mu^2 ) h 
- { \mu^4 \over 2 \lambda^2}
\ ,
\eeq
where
\beq
A^- \es
{2 \over \partial^+} \partial^\perp A^\perp  
-
{2 \over \partial^{+ \, 2} }
\left[ 
g \bar \psi \gamma^+ \psi + \kappa \partial^+ B
\right]   \ .
\eeq
Using terms equivalent to $\partial^l A^k \partial^l A^k 
- \partial^l A^k \partial^k A^l$ through integration by parts,
one obtains
\beq
\label{cHresult1}
\cH 
\es 
2 g^2 \psi_+^\dagger \psi_+ {1 \over (i \partial^+)^2 } \psi_+^\dagger \psi_+
+
2g  \psi_+^\dagger \psi_+ { 1 \over i \partial^+ } ( i\partial^\perp A^\perp - i\kappa B  )
\np
2 \psi_+^\dagger \left[ ( i\partial^\perp-gA^\perp)  \alpha^\perp + m\beta \right]  
{1 \over 2 i \partial^+} \
\left[ ( i\partial^\perp-gA^\perp)  \alpha^\perp + m\beta \right] \psi_+ 
\np
\2
A^\perp ( - \partial^{\perp \, 2} + \kappa^2 ) A^\perp 
+
\2
B ( - \partial^{\perp \, 2} + \kappa^2 )  B  
\np
\2 \, h  ( - \partial^{\perp \, 2} + 2 \mu^2 ) h 
- { \mu^4 \over 2 \lambda^2}
\ .
\eeq 
This Hamiltonian density differs from Soper's in his Eq. (23)
by the appearance of the free field $h$ with mass 
$\sqrt{2} \, \mu$. It is thus demonstrated that the 
concept of Abelian gauge symmetry understood as 
invariance under change of phase of the matter fields,
which here means fermions and scalar bosons,  leads 
in the massive limit to his result, but with an addition 
of a free field of arbitrary mass.

\subsection{ Quantization }

The FF Hamiltonian of Eq.~(\ref{Pminus}), 
$P^- = \int d^2 x^\perp dx^- \  \cH$, defined in terms of the 
density $\cH$ of Eq.~(\ref{cHresult1}), is turned into a
quantum Hamiltonian operator by a nowadays standard 
quantization procedure~\cite{KogutSoper,Yan3}. The procedure 
amounts to replacement of the classical fields by field 
operators. The field operators are constructed by imposing 
the commutation relations among their spatial Fourier 
components on the front defined by the condition $x^+=0$,
whereby the Fourier coefficients acquire the properties 
of creation and annihilation operators. The space of states 
in which the resulting Hamiltonian acts is constructed by 
acting with products of the creation operators on the 
vacuum state. The vacuum state is annihilated by all 
annihilation operators in the theory. In order to produce 
formulas for the quantum fields, it is useful to introduce 
the concept of free fields at $x^+=0$~\cite{LepageBrodsky}. 
The option of doing so only for the fields at $x^+=0$, 
{\it i.e.,} without considering their canonically conjugated momenta,  
is unique to the FF of Hamiltonian dynamics. 

The fermion constraint Eq.~(\ref{psiminus}) can be written
as~\cite{LepageBrodsky}
\beq
\label{psiminusfree}
\psi_-     \es \psi_{f-} - {g \over i \partial^+} A^\perp  \alpha^\perp \psi_+ \ , 
\eeq
where the ``free'' part is defined by
\beq
\psi_{f-} \es   {1 \over i \partial^+} \
             \left( i\partial^\perp \alpha^\perp + m\beta \right) \psi_+ \ .
\eeq
One then introduces the ``free'' fermion field $\psi_f$ by writing
\beq
\psi_f \es \psi_+ + \psi_{f-} \ .
\eeq
The vector boson field constraint Eq.~(\ref{constraintA}) 
can be written as
\beq
A^- \es A_f^-
-
{2 \over \partial^{+ \, 2} }
\left( g \bar \psi \gamma^+ \psi 
+
g'^2 \ \varphi^2 \  \kappa^{-1} \partial^+ B  
\right) \ ,
\eeq
where the ``free'' part is
\beq
A_f^- \es { 2 \over \partial^+ } \partial^\perp A^\perp \ .
\eeq
The ``free'' vector field $A_f^\mu$ is hence introduced by writing
\beq
A_f^\perp \es A^\perp \ , \\
A_f^+         \es 0               \ , \\
A_f^-         \es  { 2 \over \partial^+ } \partial^\perp A^\perp \ .
\eeq
The quantum theory is introduced by replacing the fields 
$\psi_f$, $A_f$, $B$ and $h$ by the corresponding quantum 
field operators,
\beq
\hat \psi_f
\es
\sum_{\sigma = 1}^2  \int [p]
\left[  u_{p\sigma} \hat b_{p\sigma} e^{-ipx}
      +    v_{p\sigma} \hat d^\dagger_{p\sigma} e^{ipx}
\right]_{x^+=0}
\ ,
\\
\hat A_f^\mu
\es
\sum_{\sigma =1}^2  \int [p]
\left[  \varepsilon^\mu_{p\sigma} \hat a_{p\sigma} e^{-ipx}
        +  \varepsilon^{\mu *}_{p\sigma} \hat a^\dagger_{p\sigma} e^{ipx}
\right]_{x^+=0}
\ ,
\\
\label{hatB}
\hat B
\es
\int [p]
\left[  -i \hat a_{p3} e^{-ipx}
        +  i \hat a^\dagger_{p3} e^{ipx}
\right]_{x^+=0} \ , \\
\hat h
\es
\int [p]
\left[  \hat a_{ph} e^{-ipx}
        +  \hat a^\dagger_{ph} e^{ipx}
\right]_{x^+=0} \ ,
\eeq
where $[p] = dp^+ \theta(p^+) d^2p^\perp /[2 p^+ (2\pi)^3]$,
$u_{p\sigma}$ and $v_{p\sigma}$ are the Dirac spinors, 
$\varepsilon_{p\sigma}$ are polarization four-vectors,
$\sigma$ and $\lambda$ label states of fermions and gauge 
bosons with spin projections $\pm \2$ or $\pm1$ on the 
third axis, respectively, and the creation and annihilation 
operators, denoted by $b$, $d$ and $a$, obey commutation 
or, in the case of fermions, anti-commutation rules of the form
\beq
[ \hat a_{p\lambda}, \hat a^\dagger_{q\sigma}]
\es
2 p^+ (2\pi)^3 \delta(p^+-q^+) \delta^2(p^\perp - q^\perp) \delta_{\lambda \sigma} \ ,
\eeq
with other commutators or anti-commutators equal zero. In other 
words, the Hamiltonian density $\cH$ of Eq.~(\ref{cHresult1}) 
is changed into a quantum Hamiltonian density by the substitution
\beq
\cH \rs  \cH(\psi_f, A_f, B, h) \tl 
\hat \cH \rs  \cH(\hat \psi_f, \hat A_f, \hat B, \hat h) \ ,
\eeq
and the quantum Hamiltonian operator is obtained through
\beq
P^- \rs \int d^2 x^\perp dx^- \  \cH
\tl
\hat P^- \rs 
\int d^2 x^\perp dx^- \  \hat \cH \ .
\eeq
This substitution is complemented by the normal ordering,
whereby all creation operators are moved to the left of 
all annihilation operators and the terms that result from 
commuting the operators are dropped. To simplify notation
for the quantum theory, the operator symbol ~$\hat{}$~ is 
omitted in further formulas. In the quantum FF Hamiltonian 
density written in the form
\beq
\label{cHdefinition}
\cH 
\es 
\2 \bar \psi_f \gamma^+ { - \partial^{\perp \, 2} + m^2   \over i \partial^+} \psi_f
+
\2 A^\perp ( - \partial^{\perp \, 2} + \kappa^2 ) A^\perp 
+
\2 B ( - \partial^{\perp \, 2} + \kappa^2 ) B  
\np
g  \bar \psi_f \,  / \hspace{-8pt} A_f \psi_f  
+
g  \bar \psi_f \gamma^+ \psi_f  { 1 \over i \partial^+} (- i\kappa B )
\np
\2 g^2 \bar \psi_f \gamma^+ \psi_f {1 \over (i \partial^+)^2 } \bar \psi_f \gamma^+ \psi_f
+
g^2 \bar \psi_f \, / \hspace{-8pt} A_f
{\gamma^+ \over 2 i \partial^+} \ / \hspace{-8pt} A_f  \psi_f
\np
\2 \, h  ( - \partial^{\perp \, 2} + 2 \mu^2 ) h  - { \mu^4 \over 2 \lambda^2} \ ,
\eeq
the first line describes the free FF energies of: fermions
through the field $\psi_f$; transverse gauge bosons of 
mass $\kappa$ through the field $A_f$; and gauge bosons 
of mass $\kappa$ carrying the third state of polarization 
through the field $B$. The contribution of field $B$ to 
polarization of massive gauge bosons is explained in the 
next section. The second line describes the couplings of 
fermions to the transverse bosons and longitudinal bosons. 
The third line provides the interactions that result from 
solving constraints. The first term is the FF analog of the 
IF Coulomb potential, with the inverse of $\partial^{+ \, 2}$ 
being the analog of inverse of Laplacian in the IF Gauss 
law. The second term describes the interaction due to the
instantaneous fermion propagation down the front third 
axis. The fourth line describes the FF energy of free 
quanta of field $h$, with mass $\sqrt{2} \, \mu$. Formal
discussion of the quantum theory with Hamiltonian defined
using Eq.~(\ref{cHdefinition}) can now be pursued along the 
lines indicated in Refs.~\cite{Soper,KogutSoper,Yan3} without
change, since the field $h$ is decoupled. Note, however, 
that the regularization that involves a mass parameter 
for gauge bosons alters the Bloch and Nordsieck mean-field 
approximation of Refs.~\cite{BlochNordsieck,Nordsieck}, 
because the states of modes with infinitesimally small 
$k^+$ that could be approximated using the mean field 
are blocked by the mass regularization parameter from 
being copiously produced. One has to go to the limit of 
$\kappa \to 0$ to validate the mean field approximation.

\section{ Polarization of massive gauge bosons }
\label{polarization}

The role of massive gauge boson field $\tilde B$ (we use 
the tilde notation of Sec.~\ref{TCG})  is to provide dynamical
effects of the third polarization state that a massive vector 
field can have besides the two transverse polarization states 
described by $\tilde A^\perp$ in gauge  $\tilde A^+=0$. 
One can see this by proceeding in a way analogous to 
Ref.~\cite{Soper}, except that in the case of Lagrangian 
density of Eq.~(\ref{cL}) one works in the massive limit. 
In that limit, the constraint Eq.~(\ref{constraintA}) for
$\tilde A^-$,
\beq
\label{constraintA1}
\tilde A^- \es { 2 \over \partial^+ } \partial^\perp \tilde A^\perp
-
{2 \over \partial^{+ \, 2} }
\left( g \bar {\tilde \psi} \gamma^+ \tilde \psi 
+
g'^2 \ \tilde \varphi^2 \  \kappa^{-1} \partial^+ \tilde B  
\right) \ ,
\eeq
becomes
\beq
\label{constraintA2}
\tilde A^- \es { 2 \over \partial^+ } \partial^\perp \tilde A^\perp
-
{1 \over \partial^+ } 2 \kappa \tilde B  
-
{2 \over \partial^{+ \, 2} } g \bar {\tilde \psi} \gamma^+ \tilde \psi  \ .
\eeq
When the interaction with fermions is turned off, $g \to 0$,
\beq
\label{constraintA3}
\tilde A^- \es { 2 \over \partial^+ } \partial^\perp \tilde A^\perp
-
{1 \over \partial^+ } 2 \kappa \tilde B  \ .
\eeq
To see the third polarization that corresponds to $\tilde B$, 
we change the gauge from $\tilde A^+=0$ to the one with 
$B=0$, described in Sec.~\ref{cLAtheta}. The change is 
accomplished using Eqs.~(\ref{gaugepsi}), (\ref{gaugeA}), 
(\ref{gaugevarphi}), (\ref{gaugetheta}), which in terms
of $\tilde B = - \kappa \tilde \theta$ read
\beq
\label{gaugepsi2}
\tilde \psi       \es   e^{ig f} \psi  \ , \\
\label{gaugeA2}
\tilde A^\mu   \es  A^\mu - \partial^\mu f \ , \\
\label{gaugevarphi2}
\tilde \varphi  \es \varphi \ , \\
\label{gaugetheta2}
\tilde B  \es B  - \kappa f  \ .
\eeq
Demanding $B =0$, one obtains 
$f = - \kappa^{-1} \tilde B$ and 
\beq
A^\mu   \es  \tilde A^\mu - \kappa^{-1} \partial^\mu \tilde B \ .
\eeq
This can be written as
\beq
\label{Amu}
A^\mu   \es  \tilde A^\mu + \kappa^{-1} i \partial^\mu  i \tilde B \ .
\eeq
Setting $\tilde A^\perp = 0$ and using Eqs.~(\ref{constraintA3})
and (\ref{hatB}) for free $\tilde B$ , one obtains from Eq.~(\ref{Amu}) 
that
\beq
A^\perp
\es
\int [p] \, {p^\perp \over \kappa} \,
\left[   a_{p3} e^{-ipx}
     +  a^\dagger_{p3} e^{ipx}
\right] \ , \\
A^+
\es
\int [p] \, {p^+ \over \kappa} \,
\left[  a_{p3} e^{-ipx}
     + a^\dagger_{p3} e^{ipx}
\right] \ , \\
 A^-
\es
\int [p] \, \left( {-2 \kappa \over p^+} + {p^{\perp \, 2} + \kappa^2 \over \kappa p^+ } \right) \,
\left[  a_{p3} e^{-ipx}
     + a^\dagger_{p3} e^{ipx}
\right] \ ,
\eeq
These components together form the field 
\beq
A^\mu
\es
\int [p] \, \varepsilon_{p3}^\mu \,
\left[  a_{p3} e^{-ipx}
     + a^\dagger_{p3} e^{ipx}
\right] \ . 
\eeq
where the real polarization four-vector has components 
\beq
\varepsilon_{p3} \es \left( \varepsilon_{p3}^- = {p^{\perp \, 2} - \kappa^2 \over \kappa p^+}, ~ 
                                 \varepsilon_{p3}^+ = {p^+      \over \kappa}, ~
                                 \varepsilon_{p3}^\perp = {p^\perp \over \kappa} \right) \\
                   \es  {p \over \kappa}  - \eta {\kappa \over p^+} \ ,
\eeq
and $\eta^+=\eta^\perp = 0$ while $\eta^- = 2$.
This polarization four-vector has the Minkowski 
product with the free four-momentum $p$, corresponding
to the mass $\kappa$, equal zero and its square equals $-1$. 
It complements the two transverse linear polarization four-vectors 
for the field $A$ that one obtains from the free $A^\perp$ by 
setting $A^+$ and $B$ to zero,
\beq 
\varepsilon_{p \sigma} \es
\left( \varepsilon_{p\sigma}^- = 2 p^\perp \varepsilon_\sigma^\perp / p^+, 
      ~\varepsilon_{p\sigma}^+ = 0, 
      ~\varepsilon_{p\sigma}^\perp = \varepsilon_\sigma^\perp \right) \ ,
\eeq
where $\varepsilon_\sigma^\perp = (1+\sigma,1-\sigma)/2$ with $\sigma = \pm 1$.
Together, the set of three polarization four-vectors 
correspond to massive vector quanta in agreement
with the classification of representations of the 
Poincar\'e group~\cite{Wigner}.

\section{ Conclusion }
\label{C}

The FF Hamiltonian density of Eq.~(\ref{cH1}) leads
to an interacting quantum theory that is not {\it a priori}
limited to the massive limit. Such theory involves interactions
of the field $h$ with both fields $A$ and $B$. However,
for the purpose of regularization of Abelian gauge theories
in the FF of Hamiltonian dynamics, such as FF QED, 
it appears sufficient to consider the massive limit alone.
Namely, in the RGPEP, the regularization mass parameter 
appears in the Hamiltonian interaction terms in the form
of regulating factors such as in Ref.~\cite{Gomez}, which 
for a particle of mass $\kappa$ can be replaced by 
\beq
\exp\left( - { k^{\perp 2} + \kappa^2 \over x \Delta^2 } \right) \ .
\eeq
The argument of the exponent can be written as 
$- m^2(k^\perp,x)/\Delta^2$, where $m^2$ denotes 
the particle contribution to the square of total free 
invariant mass of all particles created or annihilated by a term.
Such functions simultaneously regulate UV and small-$x$
singularities. Small-$x$ region is regulated by limiting
$x$ to values greater than about $\kappa^2/\Delta^2$.
For extremely small $\kappa$, the UV cutoff parameter
$\Delta$ provides regularization of small $x$ at extremely
small values, the smaller the larger $\Delta$. This feature
implies that a large range of small momenta $k^+$ near 
zero~\cite{Casher:1974xd,ThornRozowsky} is included 
that otherwise would be excluded by using regulating 
functions such as $\theta(x- \delta)$ with a small parameter 
$\delta$. In view of the literature on the vacuum in QFT, 
such as~\cite{Casher:1974xd,ThornRozowsky}, this is 
a welcome feature. 

The fact that the massive limit of an Abelian gauge theory 
yields also the free field $h$ of arbitrary mass does not 
produce obstacles because the field is decoupled by sending 
the independent coupling constant $\lambda$ to zero. Less
clear is the issue of quantum coupling of $B$-field quanta
to fermions. The coupling is of order $\kappa/p^+$, where 
$p^+$ is the momentum of the $B$-boson, equal to the 
momentum carried by the fermion current $g \bar \psi 
\gamma^+ \psi$, which can be arbitrarily small when the 
momentum transfer to or from the fermion approaches zero. 
Therefore, it appears that in the limit of $\kappa \to 0$ the 
$B$-bosons decouple. However, the coupling strength is 
inversely proportional to $x$ that the boson carries and this 
factor causes new small-$x$ divergence for $p^+ \ll \kappa$. 
The net effect requires precise studies that await completion.

As a final remark, we stress that the regularization of 
FF Hamiltonian dynamics of Abelian gauge theories 
using mass parameter for gauge quanta requires an
extension to the non-Abelian gauge theories in order
to become a candidate for regulating Hamiltonian 
perturbation theory in the SM. The extension requires 
that the massive limit is replaced by the condition of 
constant modulus field, which thus becomes a zero 
mode of the FF dynamics~\cite{FFreview}. The only 
dynamical quantity is the phase of the zero mode, while
its modulus becomes a regularization parameter, whose 
role is to be eventually eliminated by the RGPEP as a 
consequence of universality. Since such extension 
introduces gauge bosons of third polarization and the 
latter introduce additional singular interaction terms in 
the Hamiltonians, the following observation regarding 
the mass $\kappa$ as a regularization parameter in 
the RGPEP is in order. 

When one solves equations of the RGPEP~\cite{RGPEP}, 
the interaction terms in the resulting Hamiltonian are 
softened by vertex form factors of the form
\beq
\exp{ \left[ - t (\cM_c^2 - \cM_a^2)^2 \right] } \ ,
\eeq
where $t =s^4$ and $s$ is the size parameter for effective 
particles that plays the role of a renormalization group 
scale parameter. Symbols $\cM_c$ and $\cM_a$ denote 
total free invariant masses of the particles created and 
annihilated, respectively, by a term in the Hamiltonian 
that defines a vertex. Once one introduces the mass 
parameter $\kappa$ for quanta of fields $A$ and $B$, 
their contribution to the invariant masses in the RGPEP 
form factors is additive and in the form
\beq
{k^{\perp 2} + \kappa^2 \over x} \ .
\eeq
This form causes that the RGPEP vertex form factors 
regulate both singularities due to large $k^\perp$ and 
small-$x$ for non-zero size parameters $s$. Therefore, 
when one introduces the mass $\kappa$ in a theory 
based on gauge symmetry and uses the massive limit, 
there is no need for any separate regularization in the 
RGPEP besides the factors that automatically result form 
solving its equations.

\vskip.2in
{\bf Acknowledgment }

The author thanks Piotr Chankowski and
Davison Soper for comments.


\end{document}